# Graphene helicoid as novel nanospring

*Haifei Zhan[1,2], Yingyan Zhang[1], Chunhui Yang[1], Gang Zhang[3,*], and Yuantong Gu[2,*]*

[1]School of Computing, Engineering and Mathematics, Western Sydney University, Locked Bag 1797, Penrith NSW 2751, Australia

[2]School of Chemistry, Physics and Mechanical Engineering, Queensland University of Technology (QUT), Brisbane QLD 4001, Australia

[3]Institute of High Performance Computing, Agency for Science, Technology and Research, 1 Fusionopolis Way, Singapore 138632, Singapore

**ABSTRACT:** Advancement of nanotechnology has greatly accelerated the miniaturization of mechanical or electronic devices/components. This work proposes a new nanoscale spring – a graphene nanoribbon-based helicoid (GH) structure by using large-scale molecular dynamics simulation. It is found that the GH structure not only possesses an extraordinary high tensile deformation capability, but also exhibits unique features not accessible from traditional springs. Specifically, its yield strain increases when its inner radius is enlarged, which can exceed 1000%, and it has three elastic deformation stages including the initial delamination, stable delamination and elastic deformation. Moreover, the failure of the GH is found to be governed by the failure of graphene nanoribbon and the inner edge atoms absorb most of the tensile strain energy. Such fact leads to a constant elastic limit force (corresponding to the yield point) for all GHs. This study has provided a comprehensive understanding of the tensile behaviors of GH, which opens the avenue to design novel nanoscale springs based on 2D nanomaterials.

**KEYWORDS:** graphene nanoribbon, helicoid, spring, tensile deformation, molecular dynamics simulation

## 1. INTRODUCTION

The advancement of nanotechnology has greatly accelerated the miniaturization of electronic and mechanical devices, particularly, the nanoelectromechanical system (NEMS) [1, 2]. Benefited from the intriguing attributes, NEMS exhibits revolutionary functionalities which enable broad applications such as chemical and biological sensors [3], mass sensor or radiofrequency signal processing [4], drug delivery and imaging [5], and energy harvesting [6]. To facilitate various applications, NEMS is usually devised from the integration of multiple electronic and mechanical components [7], such as actuators, transistor, resonators, and motors. In this regard, nanospring is one of the simple building blocks for a variety of nanoscale devices. For instance, a nanomachine made from Pd nanospring is reported to show efficient propulsion in the presence of either magnetic or acoustic fields [8, 9].

Up to date, researchers have successfully synthesized nanosprings based on different types of materials, such as Pt nanowire [10], multi-walled carbon nanotube (MWCNT) [11, 12], silicon monoxide (SiO) [13], boron carbide nanowire [14], silica [15], and carbon nanocoils [16]. Meanwhile, plenty of work have been conducted to assess the electrical [17], magnetic [18], and mechanical properties of nanospring [11, 19]. For instance, through atomic force microscope measurement, the MWCNT-based nanospring is found to exhibit a nonlinear response under compression [11]. An amorphous carbon nanocoil is found to have a spring constant of 0.12 N/m in the low strain region [20]. Studies show that the $SiC@SiO_2$ coaxial nanospring has a spring constant around 6.37 N/m [21].

It is noted that previously studied nanosprings were either made from a rod structure (i.e., nanowire) or a tube structure (i.e., nanotube), analogue to the conventional spring with similar mechanical behaviours. With the emergence of diverse 2D nanomaterials, such as graphene and transition metal dichalecogenides (TMDs, e.g., $MoS_2$ and $WS_2$), a new kind of helical structure can be synthesised through the dislocation-driven growth mechanism [22-24]. Specifically, screw dislocation can create helical planes with continuous growing surface steps, leading to atomically layered spirals [24]. A very recent work shows that the $MoS_2$ spiral structure can easily carry vertical current as the topology defect in the centre connects all layers and converts the vertical transport to transverse transport in the basal plane [25]. By applying voltage to the graphene-based helicoid structure, the electrical currents flow helically and thus give rise to a very large magnetic field, which bring superior inductance [24]. These intriguing properties endow the helicoid structure with appealing functionalities for usages in nanodevices. Given the similar helicoid structures as the conventional spring, it is also of great interest to know whether they can be applied as nanosprings. Herein, by taking the graphene helicoid as a representative structure, the current work explores its tensile properties through large-scale molecular dynamics simulations. It is found that the graphene helicoid has a very large yield strain and possesses a unique tensile behaviour due to the van der Waals (vdW) interactions, distinct from the conventional spring.

## 2. METHODS

The graphene helicoid (GH) structure was constructed according to the screw dislocations as observed abundantly in annealed pyrolytic graphite [26, 27]. A representative zigzag-edged structure was chosen in this work. As shown in **Figure 1**, the GH was constructed through a single screw dislocation of graphene nanoribbon (***b***, where |***b***| = 3.4 Å) and two graphene monolayers are placed at the two ends to reduce the edge influence. Clearly, a GH structure can be defined by three parameters, i.e., the outer radius (*R*), inner radius (*r*) and turn/pitch number (*N*). The width and height of the GH can thus be calculated from $w = R - r$ and $h_{tot} = (N+2)|\boldsymbol{b}|$, respectively.

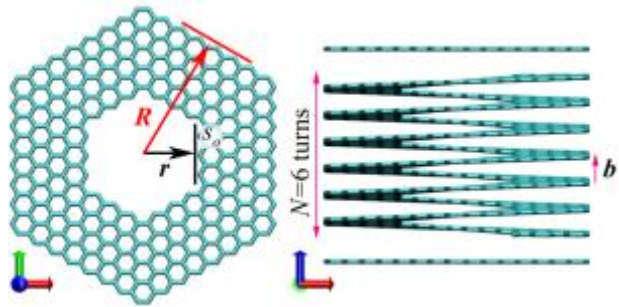

**Figure 1** Atomic structure of the GH constructed from screw dislocation, left image is top view and right image is side view.

The study was carried out by using large-scale molecular dynamics (MD) simulations. For all simulations, the widely-used adaptive intermolecular reactive empirical bond order (AIREBO) potential was employed to describe the C-C atomic interactions [28, 29]. This potential has been shown to well represent the binding energy and elastic properties of carbon materials. The cut-off distance of the AIREBO potential was chosen as 2.0 Å [30-35]. The GH structures were firstly optimized by the conjugate gradient minimization method and then equilibrated using Nosé-Hoover thermostat [36, 37] for 1 ns. Free boundary conditions were applied in all three directions. To limit the influence from the thermal fluctuations, a low temperature of 1 K was adopted for all simulations. After relaxation, a constant velocity of 0.05 Å/ps was imposed on the upper end of the GH (including the upper graphene monolayer) to realize the tensile deformation (with bottom end being fixed). Note that the loading end or fixed end contains one turn of the GH together with the monolayer graphene, i.e., the deformable or effective region has $N_{eff} = N - 2$ turns. A small time step of 0.5 fs was used for all calculations with all MD simulations being performed under the software package LAMMPS [38].

## 3. RESULTS AND DISCUSSION
### 3.1 Tensile behavior

Firstly, we acquire how the GH would behave under tensile deformation. **Figure 2a** illustrates the profile of the strain energy $\Delta E$ during the tensile deformation of GH (with $N_{eff} =$ 8, $r \sim 7.1$ Å, and $R \sim 16.33$ Å). As expected, the GH exhibits a large tensile deformation capability with a yield strain approaching to 1500%. Note that the strain is defined as $\varepsilon = h/h_o$ ( $h_o$ and $h$ are the initial and stretched heights of the deformable region, respectively). Evidently, the GH has four totally different deformation stages, denoted as stages I, II, III and IV, respectively. In stage I (see inset of Figure 2b, strain from 0 to ~ 17%), the strain energy is found to increase abruptly, which is caused by the initial delamination of two adjacent nanoribbon turns. As illustrated in Figures 2b and 2c, the initial delamination will create two surfaces and thus requires high strain energy to overcome the interlayer van der Waals (vdW) interactions. Simulation has further affirmed that this initial delaminating process is analogue to the tensile deformation of a multilayer graphene nanoribbon (in the out-of-plane direction). From Figure 2b, the strain energy curve (blue line with circle markers) for the separation of a multilayer graphene nanoribbon (with same layer number as the GH) nearly overlaps with that of the GH. Thereafter, the GH undergoes a stable delamination process and the strain energy increases when the strain increases (stage II). It is worth noting that the initial delamination may not occur in a single location and the partially delaminated regions will adhere again with further stretch (see the magenta circles in Figure 2c – 2f). Such process will cause small fluctuations to the strain energy curve, such as the abrupt energy fall in the enlarged view in inset of Figure 2a. Thereafter, delaminating process will essentially concentrate in one location (Figure 2f), and results in a smooth strain energy profile. According to Figure 2a, the strain energy is essentially a linear function of strain in stage II, indicating that the tensile force is independent of strain. After full delamination, the GH enters the third deformation stage (Figure 2g), which corresponds to the elastic deformation of the graphene nanoribbon and we observe a parabolic relationship between $\Delta E$ and $\varepsilon$ (Figure 2a). The last deformation stage (IV) corresponds to the failure of the graphene nanoribbon. Revisiting Figure 1, the inner edge of the fully delaminated GH is a zigzag edge regularly interrupted by armchair edge. In stage IV, the crack is found to initiate at the location of armchair edge and propagate along the zigzag direction, similar deformation scenario is also reported in previous studies [39, 40]. As shown in insets of Figure 2h, monoatomic chains and pentagon carbon rings are formed at the fracture region.

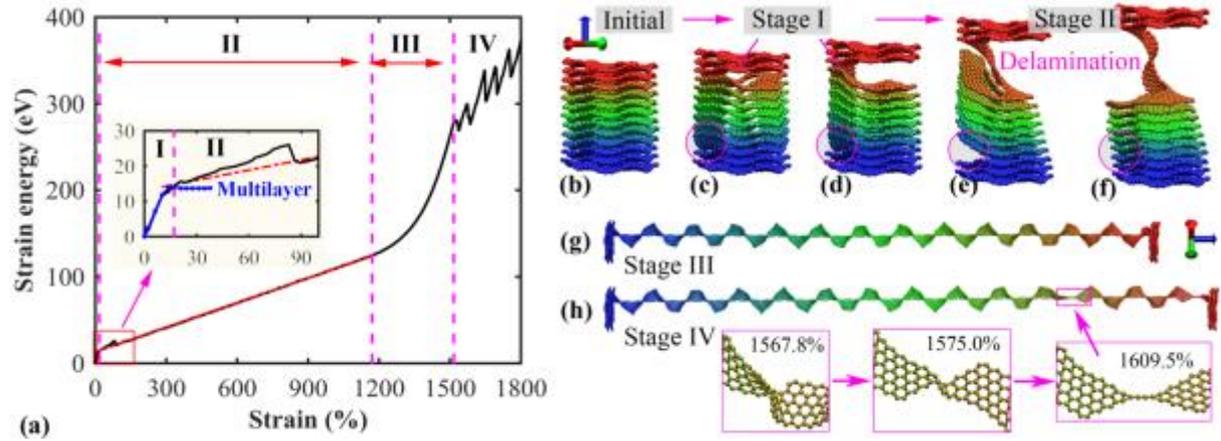

**Figure 2** (a) The strain energy of GH as a function of strain, which can be divided into four distinct regions including stage I, II, III, and IV. Inset shows the enlarged view of region I and the solid blue line (with circle marker) denotes the tensile strain energy of a multilayer graphene nanoribbon. Red dashed line represents the linear fitting line within region II. Atomic configurations showing the deformation process of the GH: (b) the initial structure; (c-d) the first deformation stage; (e-f) the delamination stage II; (g) the stretching stage III; and (h) the fracture stage IV, insets illustrate the formation of monoatomic chain at the fracture region from the strain 1567.8% to 1609.5%. Magenta circles in (c) – (f) highlight the evolution of the partially delaminated regions. Atoms are colored according to their coordinates along height direction.

Furthermore, we examine the structure of the deformed GH. It is found that the GH could fully recover to its initial state from stages I and II (see **Figure 3**a). Whereas, in stage III, the graphene nanoribbon will adhere together again, but may not fully resume to its initial structure due to the local folding/flip as resulted from its low bending rigidity (see inset in Figure 3a). Here the energy minimization is performed by removing the external tensile load from the deformed GH while keeping the other end fixed. We note that the generation of the local folding is due to the significant change of atom positions during minimization, which can be avoided if we guide the recovery process, e.g., by applying a constant velocity to compress the stretched GH. As illustrated in Figure 3a (the dashed line), the strain energy is fully released from stage III when the GH reached its initial structure. Additionally, we have also probed the impact from the tensile load rate by considering the stretching velocity ranging from 0.02 to 0.40 Å/ps, from which we observe a uniform deformation process. As evidenced from Figure 3b, the obtained strain energy profiles are nearly overlapped with each other, and the atomic configurations of the GH are found almost identical to each other at the same strain. Overall, these results have affirmed that the GH undergoes elastic deformation in these three stages, and can bear a large amount of tensile deformation.

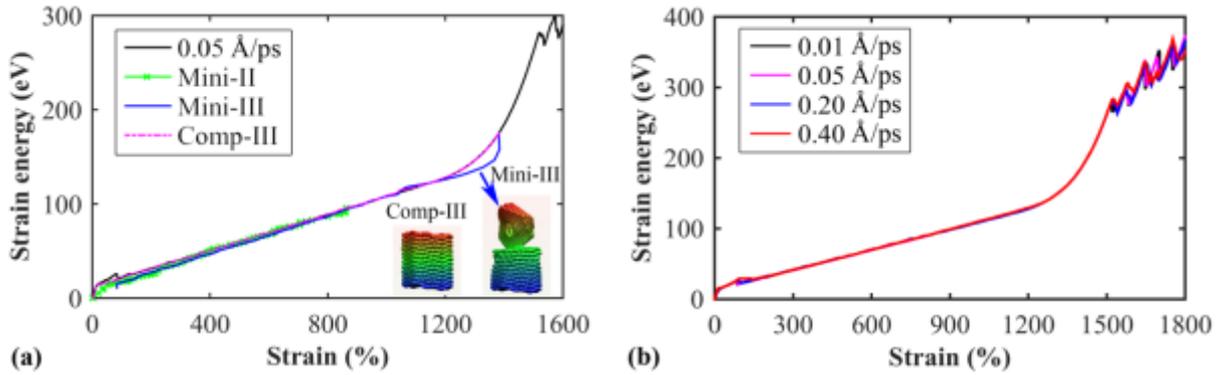

**Figure 3** (a) The strain energy of GH as a function of strain after the removal of the tensile load. Mini-II and III represent two minimization results started from stages II and III, respectively (by simply removing the tensile loads). Comp-III is the guided recovery by applying a constant compressive velocity on the top of the GH. Insets compare the atomic configurations of the GH after minimization and guided recovery from a same strain at stage III; (b) The strain energy of GH as a function of strain with different tensile velocity.

**3.2 Spring parameters**

Above tensile behavior suggests that GH could be a good candidate for nanoscale spring with a large functional zone. To explore such application, it is of great interest to parameterize the GH. As such, we revisit the change of the force during the tensile deformation for the GH with 10 total turns (see **Figure 4**). Consistent with the strain energy profile, there are four clear stages during the tensile process. Specifically, in the first stage, the force increases from zero to a local maximum value ($F_a \approx 8$ nN) to conquer the interlayer vdW interactions. For discussion convenience, we refer to this maximum force as the critical force that activates the GH spring. Thereafter, the force drops sharply to a relatively small value (~ 0.56 nN) and experiences fluctuations in the entire second stage, independent of the strain, i.e., $F_c = C$ (here, $C$ denotes constant). Such observation is understandable as the contact surfaces during the stable delamination stage (II) is much smaller compared with that in stage I, and thus leads to a much smaller force. In the third stage, the GH has been fully delaminated and the structure undergoes elastic deformation. According to Figure 4, the force shows a nonlinear relationship with the strain. Such nonlinearity is arisen from the non-uniform strain/stress distribution in the graphene nanoribbon as discussed in the following. With continuing deformation, the force reaches a new local maximum ($F_e \approx 6.04$ nN), which we refer as the elastic limit of the GH. After passing this maximum strain, bond breaking is observed, signifying the occurrence of structural failure. Overall, the mechanical performance of the GH spring is governed by the critical force ($F_a$) in stage I, the constant force ($F_c$) in stage II, the elastic limit force ($F_e$) in stage III, and their associated strain, including $\varepsilon_a$, $\varepsilon_c$ and $\varepsilon_e$ (see Figure 4).

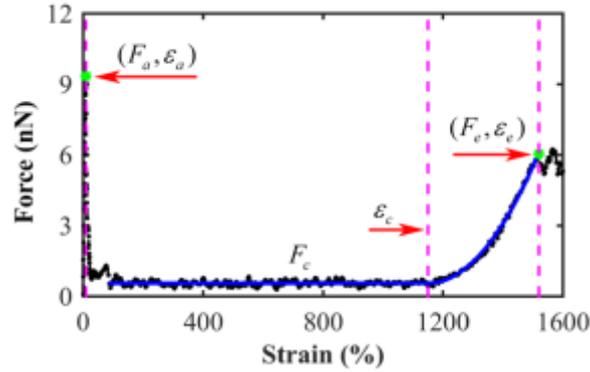

**Figure 4** A representative profile of the tensile force for the GH during stretch (with eight effective turns). The three dashed lines identify the critical parameters that govern the GH behavior.

### 3.3 Influential factors

In the following, we probe how these spring parameters are related to the geometrical parameters of the GH (i.e., $r$, $R$ and $N_{\text{eff}}$ as illustrated in Figure 1). For this purpose, three groups of GHs were tested and compared in **Figure 5**, including group one with the same $r \approx 7.1$ Å and $R \approx 16.33$ Å but changing $N_{\text{eff}}$, group two with the same $R \approx 22.72$ Å and $N_{\text{eff}} = 4$ but changing $r$, and group three with the same $r \approx 7.1$ Å and $N_{\text{eff}} = 4$ but varying $R$. Essentially, both the critical force $F_a$ and the constant force $F_c$ are determined by the vdW interactions (besides the involved bending and torsion). Thus, the larger the width is, the stronger the vdW interactions are. In other words, both $F_a$ and $F_c$ roughly follow a linear relationship with the width of the GH (see Figure 5b and 5c). However, although possessing same origin, they behave differently while altering the turn number or the width due to the different loading scenarios that the GH is experiencing. As aforementioned, the first deformation stage is analogue to the stretch of a multi-layer graphene along the thickness direction (inset of Figure 2a), i.e., a complete effective turn of the GH is under deformation in this stage. Whereas, in the second stage, the GH undergoes stable delamination process, like a peeling process and only the delamination front experiences deformation. From this perspective, it is reasonable to found that $F_a$ is much larger than $F_c$ (see Figure 5a-5c). More interesting, while $F_c$ is independent of the turn number, $F_a$ exhibits a tendency of gradual decreasing when the turn number increases (Figure 5a). It is found that $F_a$ exhibits a relatively large reduction for GH with the effective turn number $N_{\text{eff}} < \sim 30$, and saturates around 4.8 nN when $N_{\text{eff}} > \sim 30$. In this study, we have examined the initial delaminating of GH with the turn number up to 48 (i.e., $N_{\text{eff}} = 46$). One possible explanation for this phenomenon is from the interlayer energy (or van der Waals interaction) release perspective, i.e., GH with a larger turn number will experience a longer extension before entering the stable delamination process. Such a fact indicates more accumulated energy releasing during the initial delamination period, and thus it requires a smaller total force to overcome the energy barrier before the onset of stable delamination. In order to get a more comprehensive understanding of this phenomenon, further investigations are still required. Overall, it is suggested that GH with larger turn number requires a smaller force to trigger the stable delamination process.

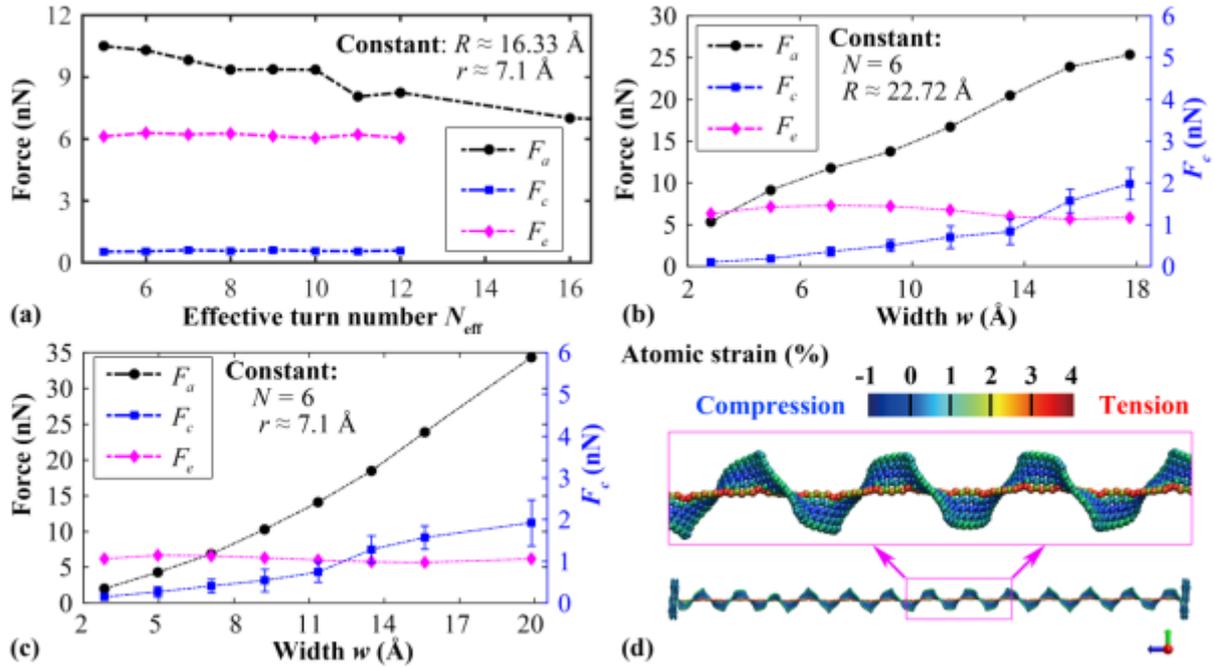

**Figure 5** The geometrical factors that influence the spring forces. (a)-(c) The critical force $F_a$, the constant force $F_c$, and the elastic limit force $F_e$ as a function of the effective turn number $N_{eff}$ and the width $w$ of the GH. In (b)-(c), the right-side $y$-axis is for $F_c$, and the errorbar represents the stand deviation of $F_c$ as estimated during the stable delamination period. (d) The atomic strain distribution for the GH at the elastic deformation stage III (at the strain of ~ 1478%). Atoms are colored according to their strain values.

Unlike $F_a$ and $F_c$, the elastic limit force $F_e$ fluctuates around a similar value for all three groups (~ 6 nN, see Figure 5a-5c), which is understandable as the elastic limit of the GH is determined by the tensile elastic limit of the graphene nanoribbon. Considering the helicoid nature of the GH, most of the tensile strain will be imposed on the inner edge atoms of the GH during the tensile deformation. For illustration purpose, we define atomic strain of the $i$th atom as $\varepsilon_i^{as} = \sum_{j=1}^{N_c} \varepsilon_{ij} / N_c$ by taking the original GH structure as the reference configuration. Here, $N_c$ is the coordination number of the $i$th atom and $\varepsilon_{ij}$ is its strain with the $j$th neighboring atom (by calculating the bond length change). As illustrated in Figure 5d, the inner edge atoms experience significant tensile strain compared to those atoms in other locations. More importantly, the atoms in the middle of the graphene nanoribbon experience compressive strain instead of tensile strain, and the magnitude of the compressive strain (~ 1%) is much smaller than that of the tensile strain (~ 4%). These observations signify that regardless of the change of turn number or width, the effective region that absorbs the stretch energy is the same in stage III and the graphene nanoribbon undergoes non-uniform deformation (with non-uniform strain distribution). This fact leads to a similar elastic limit force for GHs with different turn number or width as the carbon bond will break at the same tensile stress level.

Besides the forces, it is also of great interest to know how their corresponding strain values rely on the geometrical parameters. For the GH with changing turn number (but same inner and outer radii), $\varepsilon_a$ shares the same changing tendency with that of critical force $F_a$, i.e., it decreases with the turn number and saturates to around 5% (when $N_{\text{eff}} > \sim 34$). The largest $\varepsilon_a$ is found for the examined GH with the fewest turn number (i.e., $N_{\text{eff}} = 3$), which is about 15%. For GH with same turn number ($N_{\text{eff}} = 4$) but different $r$ or $R$ (i.e., varying width), $\varepsilon_a$ fluctuates around 11%.

In comparison, the strain limit for stable delamination ($\varepsilon_c$) is directly determined by the inner radius of the GH. Ideally, the initial height of one turn of GH is $h_o = 3.4$ Å, and the height of the fully delaminated structure can be estimated from $h_c = \sqrt{h_o^2 + 36s_o^2}$ ($s_o$ is the edge length of the inner hexagonal, see Figure 2a). Thus, the strain limit can be calculated as $\varepsilon_c = (h_c - h_o)/h_o$. Taking the GH with eight effective turns as an example, the inner edge length $s_o$ is about 7.1 Å, which yields to a delimitation limit strain around 1157% (agrees well with the results in Figure 4). Apparently, the strain limit for delamination is independent of the turn number ($N$) and the outer radius ($R$), but increases when the inner radius ($r$) increases, which is in line with the simulation results. As illustrated in **Figure 6**, for GH with the same inner radius, $\varepsilon_c$ appears nearly a constant when the outer radius (or width) increases. However, for GH with the same outer radius, $\varepsilon_c$ exhibits a significant reduction when the width increases (or the inner radius reduces). Similar as the elastic limit force, the elastic limit strain ($\varepsilon_e$) is also determined by the yield strain of the carbon bonds ($\varepsilon_{cb}$). For the fully delaminated GH with one turn, the stretchable length within the elastic limit equals $h_e = (1+\varepsilon_{cb})h_c$. That is the elastic limit strain can be calculated from $\varepsilon_e = (h_e - h_o)/h_o$, i.e., $\varepsilon_e = [(1+\varepsilon_{cb})\sqrt{h_o^2 + 36s_o^2} - h_o]/h_o$. It is evident that $\varepsilon_e$ is also independent of the turn number and the outer radius, but increases when the inner radius increases. Refer to Figure 6, a similar changing trend as observed from $\varepsilon_c$ is also found for $\varepsilon_e$, i.e., $\varepsilon_e$ maintains nearly constant for varying outer radius, but decreases significantly when the inner radius reduces.

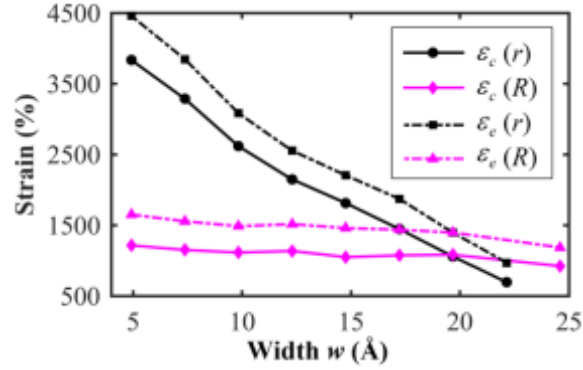

**Figure 6** The strain limit for delamination $\varepsilon_c$ and the elastic limit strain $\varepsilon_e$ as a function of the width $w$ of the GH. Here, $R$ and $r$ represent the examined group varying outer radius $R$ and varying inner radius $r$, respectively.

Finally, we probed the possible influence from hydrogen saturation at the edge. Note that different hydrogen passivation schemes at the edges of graphene nanoribbons have been demonstrated either from experiments [41] or first principle calculations [42, 43], including symmetric monohydrogenation, symmetric dehydrogenation, and asymemetric hydrogenation. In the current study, considering the π-bonds in graphene, four incline edges of the GH are considered as $sp^2$ edges and the two vertical edges are regarded as $sp^3$ edges (refer to Figure 1a), corresponding to –H and -2H full saturation conditions, respectively. The H percentage ranging from 0% to 100% were investigated by using a GH model with size as $r \approx 7.1$ Å, $R \approx 16.33$ Å, and $N = 6$. For each examined H percentage, three different cases have been considered by selecting three different random dispersion scenarios.

It is found that all examined hydrogenated GH structures exhibit a similar deformation process as observed from the pristine one. Figure 7a compares the three forces as a function of the H percentage. As it can be seen, $F_a$ and $F_e$ show a gradual reduction with the increase of H percentage (Figure 7a). Comparing with the pristine GH, $F_a$ and $F_e$ experience ~10% and ~7% reductions in the fully hydrogenated GH, respectively. Since $F_a$ is greatly influenced by the interlayer interactions, our calculations shown that H-saturation will reduce the interlayer binding energy (see Supporting Information, S1), it is thus acceptable to observe a slightly reduced $F_a$. Also, the H-saturation is supposed to weaken the C-C bonds at the edge, and thus it is not surprised that $F_e$ experiences a slight reduction with the existence of H-saturation.

Similar changing trend is also found for the corresponding strain $\varepsilon_a$, $\varepsilon_c$ and $\varepsilon_e$. As illustrated in Figure 7b, $\varepsilon_a$ decreases from ~13% to 8%, $\varepsilon_c$ decreases from ~ 1137% to ~ 900%, and $\varepsilon_c$ decreases from ~ 1517% to ~1193%, when the H percentage increases from 0% to 100%. Comparing with the pristine GH, $\varepsilon_c$ and $\varepsilon_e$ show a similar reduction amount of about 21% and $\varepsilon_a$ experience a larger reduction amount of ~36% for the fully hydrogenated GH. Being different from $F_a$ and $F_e$, the introduction of H atoms in the edge leads to a gradual increase in $F_c$. According to Figure 7a, $F_c$ exhibits a slight increase from 0.56 to 0.66 nN as the H percentage increases from 0% to 100%. According to the simulation results (see Supporting Information, S2), such increase tendency is attributed partially from the statistical estimations. For the GH with higher percentage of H, larger fluctuations of the force in stage II are observed. Overall, these results show that the H saturation exerts a small influence on the three critical forces in stages I, II, and III, but induces obvious impacts on their corresponding strain values. Additionally, we also examined the tensile behavior of the armchair-edged GH, and its size was selected as $r \approx 4.92$ Å, $R \approx 14.76$ Å, and $N = 6$. From the simulation results (see Supporting Information, S3), we found an identical deformation characteristic as observed from the zigzag-edged GH, i.e., exhibiting four similar deformation stages similar as shown in Figure 2a.

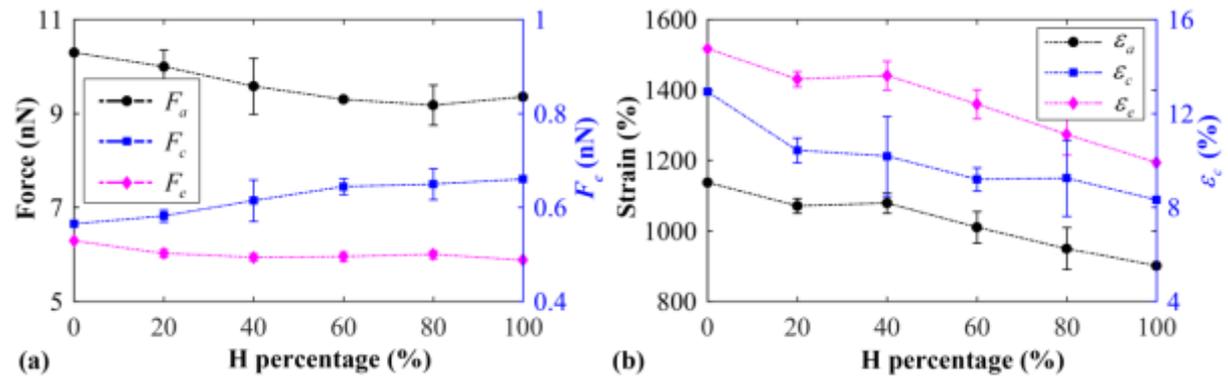

**Figure 7** The influence from H saturation: (a) the forces as a function of the H percentage, the right-side *y*-axis is for $F_c$; (b) the strain values as a function of the H percentage. The error bar represents the stand deviation as estimated from the four cases with randomly dispersed H.

## 4. CONCLUSIONS

Based on the *in silico* studies, we probed the tensile properties of the graphene helicoid (GH) made from graphene nanoribbon. It is found that the GH has a large tensile deformation capability with the yield strain exceeding 1000%. Upon the removal of the tensile load, we observe the full recovery of the structure. With this intriguing feature, we explored the application of the GH as nanoscale elastic spring. The tensile behavior of the GH can be divided into four distinct stages, including the initial delamination, stable delamination, elastic deformation and failure, respectively. Specifically, a relatively high critical force is required to trigger the stable delamination, and the tensile force maintains a constant during the stable delamination period. When the GH is fully delaminated, the force increases again until the failure of the graphene ribbon. By examining the tensile properties of various GH structures, we found that the spring parameters (including the critical forces and strains) that govern the tensile behaviors of the GH can be determined directly from the geometrical parameters (turn number and width) and the constituent graphene nanoribbon. The failure of the GH is determined by the failure of the graphene nanoribbon and the helicoid nature makes the inner edge atoms absorbs most of the tensile strain energy. This fact leads to a constant elastic limit force (corresponding to the yield point) for all GHs (around 6 nN). Additionally, it is found that the hydrogen saturation at the GH edge induces insignificant influences to its critical forces. In summary, this study has explored the tensile properties of graphene nanoribbon-based helicoid structure. The results suggest that the mechanical properties of the GH can be easily parameterized and potentially applied as a novel nanoscale elastic spring.

## AUTHOR INFORMATION

**Corresponding Author**

*E-mail: zhangg@ihpc.a-star.edu.sg; yuantong.gu@qut.edu.au

**Author Contributions**

H.Z. carried out the simulation, H.Z., Y.Z., C.Y., G.Z. and Y.G. contributed to data analysis and/or technical discussions.

**Notes**

The authors declare no competing financial interests.
## ACKNOWLEDGEMENT

Supports from the ARC Discovery Project (DP170102861) and the High Performance Computer resources provided by the Queensland University of Technology are gratefully acknowledged.

## REFERENCES

[1]    Craighead HG. Nanoelectromechanical systems. Science 2000; 290(5496): 1532-1535.

[2]     Ekinci K, Roukes M. Nanoelectromechanical systems. Review of Scientific Instruments 2005; 76: 061101.
[3]     Wang C, Madou M. From MEMS to NEMS with carbon. Biosensors and Bioelectronics 2005; 20(10): 2181-2187.
[4]     Bunch JS. NEMS: Putting a damper on nanoresonators. Nat Nano 2011; 6(6): 331-332.
[5]     Staples M, Daniel K, Cima MJ, Langer R. Application of Micro- and Nano-Electromechanical Devices to Drug Delivery. Pharmaceutical Research 2006; 23(5): 847-863.
[6]     Nafari A, Bowland CC, Sodano HA. Ultra-long vertically aligned lead titanate nanowire arrays for energy harvesting in extreme environments. Nano Energy 2017; 31: 168-173.
[7]     Kim K, Xu X, Guo J, Fan DL. Ultrahigh-speed rotating nanoelectromechanical system devices assembled from nanoscale building blocks. Nature Communications 2014; 5: 3632.
[8]     Li J, Li T, Xu T, Kiristi M, Liu W, Wu Z; et al. Magneto–Acoustic Hybrid Nanomotor. Nano Letters 2015; 15(7): 4814-4821.
[9]     Moo JGS, Mayorga-Martinez CC, Wang H, Khezri B, Teo WZ, Pumera M. Nano/Microrobots Meet Electrochemistry. Advanced Functional Materials 2017: 1604759-n/a.
[10]    Xu T, Gao W, Xu L-P, Zhang X, Wang S. Fuel-Free Synthetic Micro-/Nanomachines. Advanced Materials 2016: 1603250-n/a.
[11]    Poggi MA, Boyles JS, Bottomley LA, McFarland AW, Colton JS, Nguyen CV; et al. Measuring the Compression of a Carbon Nanospring. Nano Lett. 2004; 4(6): 1009-1016.
[12]    Deng C, Pan L, Ma H, Cui R. Electromechanical vibration of carbon nanocoils. Carbon 2015; 81: 758-766.
[13]    Zhang G, Zhao Y. Mechanical characteristics of nanoscale springs. Journal of Applied Physics 2004; 95(1): 267-271.
[14]    McIlroy D, Zhang D, Kranov Y, Norton MG. Nanosprings. Applied Physics Letters 2001; 79(10): 1540-1542.
[15]    Timalsina YP, Oriero D, Cantrell T, Prakash T, Branen J, Aston DE; et al. Characterization of a vertically aligned silica nanospring-based sensor by alternating current impedance spectroscopy. Journal of Micromechanics and Microengineering 2010; 20(9): 095005.
[16]    Cui R, Pan L, Deng C. Synthesis of carbon nanocoils on substrates made of plant fibers. Carbon 2015; 89: 47-52.
[17]    Hwang G, Hashimoto H, Bell DJ, Dong L, Nelson BJ, Schön S. Piezoresistive InGaAs/GaAs Nanosprings with Metal Connectors. Nano Letters 2009; 9(2): 554-561.
[18]    He Y, Fu J, Zhang Y, Zhao Y, Zhang L, Xia A; et al. Multilayered Si/Ni Nanosprings and Their Magnetic Properties. Small 2007; 3(1): 153-160.
[19]    Yonemura T, Suda Y, Shima H, Nakamura Y, Tanoue H, Takikawa H; et al. Real-time deformation of carbon nanocoils under axial loading. Carbon 2015; 83: 183-187.
[20]    Chen X, Zhang S, Dikin DA, Ding W, Ruoff RS, Pan L; et al. Mechanics of a Carbon Nanocoil. Nano Letters 2003; 3(9): 1299-1304.
[21]    Zhang M, Li ZJ, Zhao J, Gong L, Meng AL, Gao WD. Synthesis, growth mechanism and elastic properties of SiC@SiO2 coaxial nanospring. RSC Advances 2014; 4(85): 45095-45099.
[22]    Sarma PV, Patil PD, Barman PK, Kini RN, Shaijumon MM. Controllable growth of few-layer spiral WS2. RSC Advances 2016; 6(1): 376-382.
[23]    Zhang L, Liu K, Wong AB, Kim J, Hong X, Liu C; et al. Three-Dimensional Spirals of Atomic Layered MoS2. Nano Letters 2014; 14(11): 6418-6423.
[24]    Xu F, Yu H, Sadrzadeh A, Yakobson BI. Riemann Surfaces of Carbon as Graphene Nanosolenoids. Nano Letters 2016; 16(1): 34-39.
[25]    Ly TH, Zhao J, Kim H, Han GH, Nam H, Lee YH. Vertically conductive MoS2 spiral pyramid. Advanced Materials 2016; 28(35): 7723-7728.
[26]    Hennig G. Screw dislocations in graphite. Science 1965; 147(3659): 733-734.
[27]    Sun Y, Alemany LB, Billups W, Lu J, Yakobson BI. Structural dislocations in anthracite. The Journal of Physical Chemistry Letters 2011; 2(20): 2521-2524.
[28]    Brenner DW, Shenderova OA, Harrison JA, Stuart SJ, Ni B, Sinnott SB. A second-generation reactive empirical bond order (REBO) potential energy expression for hydrocarbons. Journal of Physics: Condensed Matter 2002; 14(4): 783.
[29]    Stuart SJ, Tutein AB, Harrison JA. A reactive potential for hydrocarbons with intermolecular interactions. The Journal of Chemical Physics 2000; 112(14): 6472-6486.
[30]    Carpenter C, Maroudas D, Ramasubramaniam A. Mechanical properties of irradiated single-layer graphene. Applied Physics Letters 2013; 103(1): 013102.
[31]    He L, Guo S, Lei J, Sha Z, Liu Z. The effect of Stone–Thrower–Wales defects on mechanical properties of graphene sheets – A molecular dynamics study. Carbon 2014; 75(0): 124-132.

[32] Zhao H, Min K, Aluru NR. Size and Chirality Dependent Elastic Properties of Graphene Nanoribbons under Uniaxial Tension. Nano Letters 2009; 9(8): 3012-3015.
[33] Zhang T, Li X, Kadkhodaei S, Gao H. Flaw Insensitive Fracture in Nanocrystalline Graphene. Nano Letters 2012; 12(9): 4605-4610.
[34] Zhao J, Wei N, Fan Z, Jiang J-W, Rabczuk T. The mechanical properties of three types of carbon allotropes. Nanotechnology 2013; 24(9): 095702.
[35] Zhan HF, Zhang G, Bell JM, Gu YT. Thermal conductivity of configurable two-dimensional carbon nanotube architecture and strain modulation. Applied Physics Letters 2014; 105: 153105.
[36] Hoover WG. Canonical dynamics: Equilibrium phase-space distributions. Physical Review A 1985; 31(3): 1695-1697.
[37] Nosé S. A unified formulation of the constant temperature molecular dynamics methods. The Journal of Chemical Physics 1984; 81: 511.
[38] Plimpton S. Fast parallel algorithms for short-range molecular dynamics. Journal of Computational Physics 1995; 117(1): 1-19.
[39] Xia K, Zhan H, Hu Da, Gu Y. Failure mechanism of monolayer graphene under hypervelocity impact of spherical projectile. Scientific Reports 2016; 6: 33139.
[40] Yin H, Qi HJ, Fan F, Zhu T, Wang B, Wei Y. Griffith Criterion for Brittle Fracture in Graphene. Nano Letters 2015.
[41] Fujii S, Ziatdinov M, Ohtsuka M, Kusakabe K, Kiguchi M, Enoki T. Role of edge geometry and chemistry in the electronic properties of graphene nanostructures. Faraday Discussions 2014; 173(0): 173-199.
[42] Deng X, Zhang Z, Tang G, Fan Z, Zhu H, Yang C. Edge contact dependent spin transport for n-type doping zigzag-graphene with asymmetric edge hydrogenation. Scientific Reports 2014; 4: 4038.
[43] Wagner P, Ewels CP, Adjizian J-J, Magaud L, Pochet P, Roche S; et al. Band Gap Engineering via Edge-Functionalization of Graphene Nanoribbons. The Journal of Physical Chemistry C 2013; 117(50): 26790-26796.